# Sustainable and resilient strategies for touristic cities against COVID-19: an agent-based approach

Marco D'Orazio[1], Gabriele Bernardini[1], Enrico Quagliarini[1,*]

[1] Department of Construction, Civil Engineering and Architecture, Università Politecnica delle Marche, via di Brecce Bianche 60131 Ancona

CORRESPONDING AUTHOR: Enrico Quagliarini, mail: e.quagliarini@staff.univpm.it - phone: +39 071 220 4248, fax: +39 071 220 4582

**Abstract.**
Touristic cities will suffer from COVID-19 emergency because of its economic impact on their communities. The first emergency phases involved a wide closure of such areas to support "social distancing" measures (i.e. travels limitation; lockdown of (over)crowd-prone activities). In the second phase, individual's risk-mitigation strategies (facial masks) could be properly linked to "social distancing" to ensure re-opening touristic cities to visitors. Simulation tools could support the effectiveness evaluation of risk-mitigation measures to look for an economic and social optimum for activities restarting. This work modifies an existing Agent-Based Model to estimate the virus spreading in touristic areas, including tourists and residents' behaviours, movement and virus effects on them according to a probabilistic approach. Consolidated proximity-based and exposure-time-based contagion spreading rules are included according to international health organizations and previous calibration through experimental data. Effects of tourists' capacity (as "social distancing"-based measure) and other strategies (i.e. facial mask implementation) are evaluated depending on virus-related conditions (i.e. initial infector percentages). An idealized scenario representing a significant case study has been analysed to demonstrate the tool capabilities and compare the effectiveness of those solutions. Results show that "social distancing" seems to be more effective at the highest infectors' rates, although represents an extreme measure with important economic effects. This measure loses its full effectiveness (on the community) as the infectors' rate decreases and individuals' protection measures become predominant (facial masks). The model could be integrated to consider other recurring issues on tourist-related fruition and schedule of urban spaces and facilities (e.g. cultural/leisure buildings).

## 1. Introduction

The smart adaptation of cities against different risks is one of the key challenges for their sustainability and the resilience of the hosted communities (C. Chen et al., 2020; Ribeiro and Pena Jardim Gonçalves, 2019). Urban areas involved by tourists' flows represent a particular application context for such resilience issues, because of the complexity between economic, social (including relationships between tourists' and residents' needs) and organizational tasks, especially in those scenarios in which seasonal tourism is a training element for the community (Feleki et al., 2018; Qie and Rong, 2016; Stanganelli et al., 2020). Due to such aspects, touristic areas are generally more susceptible to disaster effects than the other urban contexts (Aznar-Crespo et al., 2020; Rosselló et al., 2020).

One of the fundamental short-terms challenges for such touristic urban areas is surely represented by the COVID-19 emergency (Gössling et al., 2020; Iacus et al., 2020; Jamal and Budke, 2020; Nicola et al., 2020). In fact, they represent a significant scenario for the contagion spreading, essentially because the possibility of interactions among the individuals (in a direct or indirect way) is boosted by possible significant conditions in (Chakraborty and Maity, 2020; Yang et al., 2020): 1) interactions between visitors and residents (mainly, in public areas, accommodation, other tourist facilities and leisure buildings) with the possibility to "import" positive cases into the touristic areas (towards local outbreaks) or "export" them; 2) crowd levels, which cannot be always managed by the stakeholders (e.g. crowd in outdoor public spaces), thus amplifying the transmission probabilities. The same risks can be connected to international, national and local tourists' flows.

Such areas suffered (and are still suffering) the immediate counteract pandemics measures concerning "lockdown" solutions (i.e. restricted mobility and travels, "social distancing"), adopted by most of the Countries, thus proposing a blockage of touristic flows in the "first phase" of the emergency (Anderson et al., 2020; Bruinen de Bruin et al., 2020; Gössling et al., 2020; Hellewell et al., 2020; Iacus et al., 2020; Jamal and Budke, 2020; Prem et al., 2020; Yang et al., 2020). Figure 1 shows how such strategies have been generally and gradually reduced the number of active cases[1].

---

[1] e.g. for international statistics, see https://shiny.rstudio.com/gallery/covid19-tracker.html (in Italian - last access: 12/05/2020)

The return to "business as usual" should consider how differences among different Countries still exist (as well as among areas in the same Country) because of the initial conditions of the contagion. To the date this paper is written (early May 2020), considering Italy, which was one of the most COVID-19 affected Countries in the "first" emergency phase[2], the overall percentage of active cases over the population for the whole National territory is equal to about 0.16%, while the Lombardia Region (where many initial outbreaks happened (De Natale et al., 2020)) is still experiencing values over 0.30% (average values from the 28th of April to the 12th of May 2020). Local conditions can raise the current percentage up to values over 1% (e.g. compare to the data from Republic of San Marino, which is placed inside the mainland Italy and near to areas with a significant contagion spreading in the first outbreaks)[1]. Hence, mitigation measures should be balanced with respect to the number of active cases as well as to the benefits for the overall society ((Bin et al., 2020; Ferguson et al., 2020), to ensure that local areas characterized by higher active percentage cases and consistent asymptomatic ratio within the population could not lead to severe second peaks in the contagion (Anderson et al., 2020; Feng et al., 2020; Mizumoto et al., 2020; Prem et al., 2020; Roda et al., 2020).

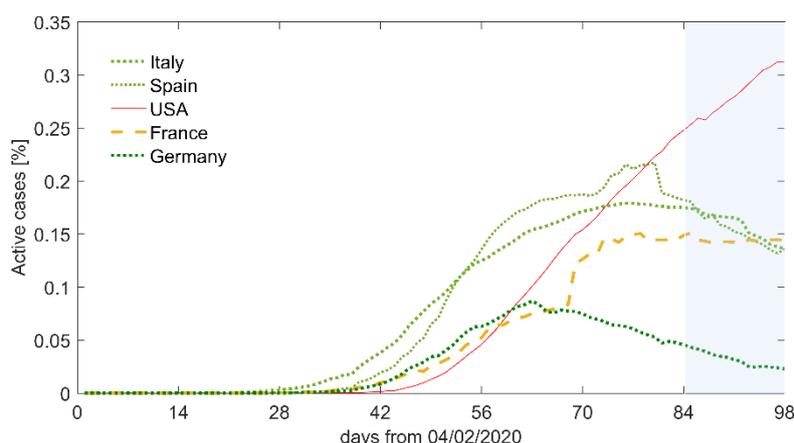

*Figure 1. Trend of the COVID-19 active cases (percentages on the population) per Country and over the time, according to international databases[1], starting from the 4h of February to 12th of May 2020, by distinguishing countries in which the percentage of the active cases is decreasing (green dotted curves), stable (yellow dashed curve) or still increasing (red continuous curves). The final blue area evidences the trend in the last 14 days before the simulation are performed to trace the main current conditions of the contagion spreading.*

Since current consolidated rules for contagion spreading are based on a coupled proximity-based and exposure time-based mode of transmission (Bourouiba, 2020; Ferguson et al., 2020; Hamid et al., 2020; Yang et al., 2020)[3], quick-to-apply non-pharmaceutical interventions are mainly aimed at (Barbieri and Darnis, 2020; Bruinen de Bruin et al., 2020; Carlos Rubio-Romero et al., 2020; Feng et al., 2020; Howard et al., 2020; Pradhan et al., 2020; Prem et al., 2020; Servick, 2020; Yang et al., 2020): 1) increasing social distancing solutions, utilizing restricted access especially in closed environments and limitations to travels (for both residents and visitors); 2) use of respiratory protective devices (facial masks) to mitigate the effects of not respected distances between the individuals. Besides, the control and tracking of COVID-19 cases could be implemented to "isolate" them or limiting possible contact with susceptible people (Kumar et al., 2020; Madurai Elavarasan and Pugazhendhi, 2020), by improving the effectiveness of the quick-to-apply measures. Nevertheless, such kind of strategies could imply a higher level of complexity for the whole urban (or even territorial) scale (e.g. monitoring the cases over time and space by means of individual tracking solutions and health checks also for asymptomatic individuals) and building scale (i.e. access control strategies by localized and rapid health checks, e.g. fever detection at the building entrance, which need widespread trained staff and specific equipment).

As for other kinds of resilience-related issues in the urban areas (C. Chen et al., 2020; Miller, 2015; Ribeiro and Pena Jardim Gonçalves, 2019), decision-makers will can select acceptable solutions according to a holistic approach which should consider: 1) final users (including tourists), to restore "normal" fruition conditions as well as possible, and their reciprocal interactions within the Built Environment of our cities (e.g. movement, activities); 2) specific stakeholders to bring together economic aspects (e.g. maximization of tourist capacity in safe conditions) and operational tasks (towards quick-to-apply and "cheaper" solutions); 3) the interactions among them, by using their representation in the considered scenarios.

---

In this view, it is necessary to provide support tools for the decision-makers, to evaluate the effective impact of each measure and their combination, with regard to the complex interaction system regulating the pandemic dynamics in the considered scenario (D'Orazio et al., 2020; Ronchi and Lovreglio, 2020).

Simulation tools can increase the awareness of decision-makers in understanding the impact of mitigation solutions on the virus spreading depending on possible scenario conditions (Bin et al., 2020; Ronchi and Lovreglio, 2020; Zhang et al., 2018). The contribute of simulation models in developing and testing strategies for risk reduction has been widely evidenced in many different cases concerning individuals' safety at the different scales of the Built Environment (both involving single buildings and urban scale), and especially in all the cases in which individuals' behaviours (including motion issue) can affect the safety levels for the individuals and the whole community (i.e. emergency evacuation modelling) (Bernardini et al., 2017; Y. Chen et al., 2020; D'Orazio et al., 2014; Lovreglio et al., 2020).

In a pandemic-risk related context, decision-makers can be supported by macroscopic Susceptible-Infectious-Recovered/Removed (SIR) and Susceptible-Exposed-Infectious-Recovered/Removed (SEIR) models (Banos et al., 2015; Hethcote, 1989), which can include general rules for moving individuals within the overall population to take into account the dynamics due to mobility issues (Boccara and Cheong, 1992). SIR/SEIR-based models have been developed also for the COVID-19 emergency, e.g. (Feng et al., 2020; Lopez and Rodo, 2020; Prem et al., 2020; Roda et al., 2020). These epidemiological models can supply decision-makers with prediction data at large scales (territorial/national) which include the effects of different levels of non-pharmaceutical interventions. Nevertheless, one of their main limits is related to the scarce level of representation of specific patterns in individuals' mobility behaviours and interactions within the Built Environment, especially while investigating smaller areas (e.g. parts of a city; single building or group of buildings; complex facilities and environment, including transportations) (Boccara and Cheong, 1992; Goscé et al., 2015; Ronchi and Lovreglio, 2020; Zhang et al., 2018). Efforts in creating microscopic models for the COVID-19 spreading within the users in the Built Environment have been performed, to take into account behavioural dynamics in spaces use (D'Orazio et al., 2020; Fang et al., 2020; Ronchi and Lovreglio, 2020), thus leading towards better awareness-based support tools for decision-makers in urban areas or single buildings. In general terms, they adopted the consolidated proximity-based and exposure-time-based rules for the transmission probability, to estimate the impact of all direct and indirect contagion effects between individuals placed at a close distance (Fang et al., 2020), but different transmission modes have been included by some approaches (Ronchi and Lovreglio, 2020). In particular, this research group recently developed and tested a proximity-based and exposure-time-based simulator according to an Agent-Based Modelling (ABM) approach, to estimate the contagion spreading in public buildings (D'Orazio et al., 2020). It includes the possibility to consider both the movement of people and the implementation of different risk-mitigation strategies (i.e. facial masks, social distancing, and access control strategies), according to a probabilistic approach. The model has been calibrated according to experimental data to provide reliable outcomes for the considered conditions. Meanwhile, the ABM approach ensures the possibility to modify the behaviours of the simulated individuals to easily adapt the simulator to other contexts in which the individuals' motion is relevant for the contagion spreading, such as the touristic cities (Banos et al., 2015).

This study adopts this simulation approach to estimate the virus spreading in tourist urban areas, depending on different surrounding input scenarios such as density conditions (including the tourist-residents ratio), tourists' characterization (e.g. holiday permanence, activities and movements in the urban areas), pandemic conditions (i.e. the initial percentage of active cases) and the implementation of risk-mitigation strategies (i.e. social distancing, facial mask use by the simulated population). To this end, modifications to the original model have been provided to ensure the application to touristic urban areas, while sensitivity analysis (Sobol', 2001) is adopted to estimate the impact of each input variable on the final results. According to a conservative approach in the quantification of infected people during the time, the epidemiological model has been extended to the whole simulation environment, thus not considering the possibility that outdoor conditions could mitigate the contagion probability. The model has been applied to a significant case study (a part of a touristic coastal city in Italy) to demonstrate its capabilities in evaluating the impact of different mitigation strategies on the infected people's number.

## 2. Phases, model description and methods

This work is divided into the following phases:
1) selection of modelling approach by modifying an existing calibrated simulation tool (D'Orazio et al., 2020) (see Section 2.1);
2) selection of a significant application case study to perform the simulation according to a sensitivity-based approach which allows refining the adopted variables within the model (see Section 2.2);
3) analysis of the results for the case study application, to evidence the effects of the main considered variables in the view of the sensitivity-based model refining (i.e.: tourists' capacity, facial masks implementation by the population, initial infector percentages) (see Section 2.3).

### 2.1. Modelling approach

The ABM model adopted in this work is based on the one proposed by (D'Orazio et al., 2020) and jointly represents the contagion spreading and the movement of simulated individuals in the considered touristic urban area. The model adopts a probabilistic approach for simulating both the aspects and has been implemented in a simulation software through the NetLogo platform (Wilensky, 1999). An R script (R version 3.6.3[4]) is implemented to perform an adequate number of simulation according to previous research approaches on epidemiologic researches (Banos et al., 2015).

Concerning the epidemic rules, the proximity-based contagion spreading approach is implemented according to previous works on consolidated COVID-19 epidemic rules[3] (Banos et al., 2015; Fang et al., 2020; Yang et al., 2020). In the model, the probability that a susceptible individual *i* can be infected by an infector *j* when they are placed at a distance equal or lower than 2m within the simulated environment at the current time depends on the linear combination between:

1. the current incubation time of *j*. The contagion probability will be maximized when the maximum incubation time is reached (according to a conservative approach, 5.1 days, which refers to the median incubation time, and the lower bound of the confidence interval, given by previous work(Lauer et al., 2020));
2. the exposure time, which is the time spent in contact by two individuals (maximized for a contact of 15 minutes);
3. the mask filter protection respectively adopted by *i* and *j* (from 0, which implies "No mask" conditions to 1, which corresponds to maximum protection level, e.g. FFP3 according to EN 149:2009).

These epidemic rules can represent all the direct and indirect contacts that can happen between the simulated agents. At the start of the simulation, a certain *initial infector %* is defined by the user. In the next steps, the contagion probability is calculated according to the aforementioned criterion. As in the original model, once a susceptible individual is infected, he/she will become an infector after a "delay" period, which is considered to be equal to 1 day (Lauer et al., 2020). The infected agents who are not-asymptomatic can exit the simulation (can "die") when the fever onset time (from the contagion) is reached. This time is considered as a variable between one day and the considered incubation time (5.1 days) (Lauer et al., 2020).

The touristic urban area (called "*world*") that hosts the agents is modelled as "a unique layer whose total area depends on the gross one of the space to be simulated", according to the original model. Hence, the world gross area can be calculated as the sum between the accommodation areas and the other areas where people can spend time during the holidays (e.g. beaches, parks, city centre avenues, shopping centres, restaurants and so on). The *world* is divided into patches according to a 1:1 scaled representation of the urban areas (1 patch = 1 m).

The original model has been modified to take into account the possibility to represent two main agents' typologies: *tourist* and *residents*. At the start of the simulation, the *tourists* and the *residents* are generated within specific areas of the *world* (which are expressed in terms of percentage of the overall world, by respectively defining the k*tourist* and k*resident* percentage values[5]). An *initial-distance* of generation has been introduced to consider "social distancing" behaviours between the individuals from the beginning of the simulation. However, when the agents' density does not allow the observance of the imposed *initial-distance*, the *initial-distance* is equal to the maximum achievable one. By this way, the agents are uniformly generated as well as possible within the *world*.

During the simulation, the tourists remain within the *world* for the holiday period (*mean-permanence* variable), and will be generated again when the holiday period will be completed, to simulate the departures/arrivals of visitors. On the contrary, there are no new births and travel into or out of the simulated population for the residents. In this sense, *residents* can only "die", that means exiting from the simulation (people who spontaneously leave/not enter the urban space due to their health conditions), while infected tourists can be "re-generated" because of the above. According to the average duration of holidays in Italy from recent national statistics[6], the overall simulated time is set to 3 days (288 steps according to the adopted time discretization, see later). This can allow a rapid tourist "renovation", thus leading towards more critical contagion conditions within the overall population.

Movements rules for *tourists* and *residents* depend on the specific time of the day in which they are performed, by considering a time discretization of 15 minutes (1 simulation step), according to the exposure time. Depending on the moment of the day, each agent can be involved in:

• morning/afternoon/evening activities: randomly moving in the city areas by the *movement-at-breaks* value, to evidence an equal probability of interacting with any other person within the *world*;
• lunch/dinner: moving near the initial generation position by trying to maintain the *initial-distance*;
• night sleep: remaining at the initial generation position.

The whole day time is represented by considering about 8 hours for night-time for sleeping. Every 96 steps (corresponding to 1 day), the activities restart again.

---

## 2.2. Case study definition and sensitivity analysis criteria

The considered case study involves a typical coastal touristic city characterized by a high density of tourists during the summer holiday. In this sense, Italian cities of the Adriatic Coast (the so-called "Riviera Romagnola", placed in the Emilia Romagna region) represent a significant application scenario. In a typical city of this context, most of the tourist accommodations are generally represented by hotels placed in the city centre areas, close to the beaches, with an overall building density which can reach over 6000 persons per squared kilometre and a ratio between tourists and residents that can be about over 10 to 1[7]. According to the criteria for dimensioning rooms and collective spaces (e.g. spaces used as restaurants, halls and so on) in hotels for the Italian national standards[8], a typical hotel density can range from 0.1 to 0.2 pp/m², by considering an average number of about 160 tourists hosted in each hotel7. According to the criteria for beach resorts organization in the application context[9], an overall density of 0.2 can be considered for the spaces used by the tourist along the beaches.

Table 1 resumes the other variables adopted in this study, while Table 2 traces the values of the constant parameters.

In view of the above, the considered case study involves about 10 hotels by considering a part of the touristic city centre scenario described above, by hosting a maximum number of individuals $N$ equal to 1600 persons over an overall area of about 20000m² (represented by a square *world* with a side of 145 patches). In each simulation, a minimum tourist capacity of 20% is defined for the minimum $N$ value.

The maximum value for the *initial infector %* is arbitrarily chosen to recreate a possible critical scenario for a "second phase" in the COVID-19 emergency basing on current national[2] and international[1] data on the contagion spreading (i.e. about 10 times the maximum number of active cases from 28th of April to 11th of May 2020, to include possible significant differences between undetected and detected CODIV-19 cases). The *initial-distance* is set up to take into account the possibility of implementation of "social distancing" strategies, by allowing a general maximum distance between individuals over the proximity distance limit for the contagion probability calculation, equal to 2m. The maximum *mean-permanence* value refers to the maximum incubation time according to consolidated international organization sources[3]. *ktourist* and *kresident* are considered variables between 0 and 1 to simulate different levels of interactions between the two agents' typologies also in respect to the accommodation type (i.e. different levels of contacts among the accommodation staff and the hosts), and the tourist-fraction is considered as variable between 0 and 100% so as to consider differences in the "die" behaviours considered in the model. Finally, constant parameters in Table 2 are chosen according to the model calibration process (D'Orazio et al., 2020) according to consolidated sources of the COVID-19 contagion, to have a consistent scaling of the contagion phenomenon in view of a 24-hours-extended simulation of the considered scenario.

The considered scenario is involved in sensitivity analysis thanking the R script which implements the NLRX package of "R statistics" programming language (Salecker et al., 2019). Variance-based decomposition methodology by Sobol (Sobol′, 2001) is used to this end according to the adoption of the sobol2007 function of "R statistics" (Saltelli et al., 2010, 2007). For any considered stochastic input in the simulation, two indexes are calculated (Saltelli et al., 2010, 2007):

1. the total index (Sobol Total index - STi) represents the effects to the output variance (including those related to interactions with other inputs). The higher the STi, the most influential the considered input on the result;
2. the first-order index (Sobol First-order index- SFi) measures the main contribution of the considered input to the variance of the output.

We performed two sets of 27000 runs. The first set considers all the variables defined in Table 1, which also describes the selected Probability Density Functions (PDFs). Then, the variables with a STi<0.05 are reasonably considered as not influential on the model output variance (Saltelli et al., 2007). Hence, in the second simulation set, they were considered as constant parameters (equal to the mean of the uniform distribution). Such simulations are analyzed to define the impact of different parameters and risk-mitigation strategies in the considered scenario, according to the criteria exposed in Section 2.3.

| Parameter (unit of measure) | Min | Max | PDFs |
|---|---|---|---|
| N (pp) | 320 | 1600 | Uniform |
| Initial infectors % (%) | 0 | 10 | Uniform |
| Mask wearing % (-) | 0 | 1 | Uniform |
| Mask filter (-) | 0 | 1 | Uniform |
| Movement at breaks (m, equal to patches) | 1 | 100 | Uniform |
| Initial-distance (m, equal to patches) | 1 | 3 | Uniform |

---

[7] e.g. compare to the data from Cattolica (RN, Italy): for general data
https://ugeo.urbistat.com/AdminStat/it/it/demografia/dati-sintesi/cattolica/99002/4 ; for tourist information
https://bit.ly/3dDE4Vy (in Italian - last access: 10/05/2020)
[8] https://www.gazzettaufficiale.it/eli/id/2009/02/11/09A01326/sg (in Italian - last access: 10/05/2020)
[9] https://imprese.regione.emilia-romagna.it/turismo/temi/demanio-marittimo-turistico-ricreativo-e-portuale/ordinanza-balneare-1-2018 (in Italian - last access: 10/05/2020)

| Mean-permanence (days) | 1 | 14 | Uniform |
|---|---|---|---|
| Tourist-fraction (%) | 0 | 100 | Uniform |
| kresident (-) | 0 | 1 | Uniform |
| ktourist (-) | 0 | 1 | Uniform |

*Table 1. Variables characterization for the simulations (first simulation set in the sensitivity analysis)*

| Parameter | Value | Source |
|---|---|---|
| $p_{imm}$ | 0 % | no evidence that immune people can exist |
| asymptomatic ratio | 20% | as for the original model and the calibration tests, chosen as the upper bound of the confidence interval of estimated asymptomatic proportion (among all infected cases) in previous experimental conditions (Mizumoto et al., 2020) |
| delay | 96 | equal to 1 day (96 steps of 15 minutes within 24 hours of simulation) to be shorter than the time to fever onset by the 2.5% of infected persons (Lauer et al., 2020). The value is scaled from the original model calibration setup. |
| Iinc | 512 | according to a conservative approach, corresponding to the median incubation time, which is about 5.1 days as in previous consolidated data (Lauer et al., 2020). The value is scaled from the original model calibration setup. |
| Ifev | 256 | the average value corresponds to the minimum time to fever onset by the 2.5% of infected persons (Lauer et al., 2020). A standard deviation is associated with range the individual's value from 0 to 512 steps. The value is scaled from the original model calibration setup. |

*Table 2. Constant parameters characterization for all the simulations. References to values from the original model setup are wider discussed in (D'Orazio et al., 2020).*

## 2.3. Criteria for results analysis

The results concerning the second simulations set used in the sensitivity analysis are used to compare the effects of the main independent variables affecting the contagion spreading.

According to previous simulation studies on COVID-19 spreading in the Built Environment (D'Orazio et al., 2020), the final infected people percentage *dI* (%) is considered to trace the contagion spreading at the end of the simulation and evaluate the variables conditions affecting the final result. *dI* depends on the ratio between the final number of susceptible people (not infected) *Sf* (pp) and the initial number of susceptible people (not initially infected) *Si* (pp), as shown by Equation 1:

$$dI = \left[1 - \frac{S_f}{S_{init}}\right] \% \quad (2)$$

*dI* allows comparing different conditions in terms of *initial infector %* as well as of *N* (including the possibility of a reduction during the time because of "die" behaviours). Higher *dI*, higher the contagion spreading among the simulated population. According to the adopted probabilistic approach, different *dI* will be produced for the same combinations of the variables. Hence, *dI* distributions (by using distribution percentiles and boxplot representation, by excluding the outliers) are assessed in respect to the input variables combinations. Furthermore, *dI* acceptability limits for the solution effectiveness are provided according to *dI*=5% and *dI*=25%, according to a percentile-based analysis of the output values. The application of these limits allows filtering the specific input combinations that ensure the possibility to have the related *dI* values respected, thus evidencing the impact of the considered variables.

From this point of view, *dI* outputs are discussed according to the current solutions in contagion spreading reduction (D'Orazio et al., 2020; Yang et al., 2020; Zhai, 2020) as well as the current contagion spreading conditions (e.g. the number of active cases in reference to experimental data[1,2]).

Facial masks effects are assessed with respect to the combination between the *mask filter* and *mask wearing %*. To have a first synthetic and quick evaluation on such strategy, the *dI* distribution is assessed by coupling *mask filter* and *mask wearing %* (*mask wearing %\*mask filter*), according to homogeneous classes with steps of 10%. Furthermore, 3 *mask filter* classes with similar dimension in terms of uniform input distribution (see Table 1) have been considered to assess the impact of different kind of implemented masks:

- respiratory protective devices, representing FFP1, FFP2 and FFP3 masks according to the EN 149:2009 (Carlos Rubio-Romero et al., 2020), are considered in the *mask filter* range from 0.80 to 1.00;

- surgical masks are considered within the *mask filter* range 0.58 to 0.83, which is placed inside the limits for classifying surgical mask efficiency according to the NIOSH NaCl method proposed by (Rengasamy et al., 2017);
- non-standard protection solutions (e.g. home-made and non-certified protections) (Carlos Rubio-Romero et al., 2020) can reasonably consider the first quartile in *mask filters* values, thus ranging from 0.00 to 0.25 as for the application of the reference model.

*N* conditions are organized with respect to the urban population density, and so the tourist capacity (i.e. for tourist-fraction equal to 100%). For each simulation, the normalized occupants' density *Docc* (-) for the overall simulated urban environment (the whole *world*) is calculated as the ratio between the current *N* value and the maximum one. Hence, according to the *N* distribution limits in Table 1, *Docc* varies from 0.2 (for *N*=320 people) to 1 (maximum occupancy of the urban areas for *N*=1600 people).

Finally, different classes of *initial infector %* are considered to take into account different input conditions about the current situation of the contagion within the population[2]. Data on active COVID-19 cases for the international and national scenario from the 28[th] of April to the 12[th] of May 2020 have been considered (time period which corresponds to the maximum incubation time before the simulations) to create the following limit conditions:

- current average active cases conditions in Italy: *initial infector %*=0.15%;
- current average active cases conditions in the worst Italian region: *initial infector %*=0.30% (referred to the Lombardia region);
- current maximum active cases conditions: *initial infector %*=1.40%, for all the monitored Countries. This data corresponds to the situation of San Marino Republic on 7[th] of May 2020 (it is worth to notice that San Marino is an interesting maximum reference data in respect to the closeness with the considered application context).

Additional limits for 3% and 5% are also proposed to evidence the possibility of critical contagion spreading conditions, also according to the previous simulation models application. The *initial infector %* values are then organized and discussed by referring to classes according to such limits.

Since the *initial infector %* highly affect the possibility to reach widespread contagion conditions within the Built Environment during the simulation time (Bin et al., 2020; D'Orazio et al., 2020; Hellewell et al., 2020; Prem et al., 2020), the difference in infectors percentage *dINF* (-) is additionally evaluated to evidence if particular additional conditions could suggest that the contagion conditions will not make worse at the end of the simulation (*dINF*>0), according to equation 2:

$$dINF = initial\ infector\ \% - dI\ (\%)\ (2)$$

In fact, lower the initial number of infectors (e.g. because of severe active cases control strategies), higher the possibility to maintain *dI* under the acceptability threshold (especially for low occupants' densities conditions), lower the absolute *dINF* value.

Finally, results are organized in the view of defining simple rules to estimate the impact of measures combination. According to previous simulation works suggestions (D'Orazio et al., 2020), *mask filter-Docc* pairs are correlated by filtering the values which allow maintaining *dI*≤5%, by additionally investigating the impact *mask wearing %* and *initial infector %* variations. In particular, the assessment is performed to evidence how the initial contagion conditions could alter the efficiency of the predicted measures.

## 3. Results

### 3.1. Sensitivity analysis and robustness check

Figure 2 displays the total order sensitivity indices (STi) and the first-order sensitivity indices (SFi) for the first simulation set. The first-order index represents the main effect contribution of each input factor to the variance of the output. The total order index accounts for the total contribution to the output variation due to factor *Xi*, i.e. its first-order effect plus all higher-order effects due to interactions.

The Total order sensitivity indices (STi) suggests how the main source of results' uncertainty is *N*. The secondary source of results' uncertainty is represented by *Initial infectors %* while the importance of individual protection measures is confirmed by the STi value assigned to *mask filter* and *mask wearing %*, thus confirming the previous model application (D'Orazio et al., 2020). The characterization of tourists' conditions is another significant source of uncertainty, as demonstrated by the STi value of the *tourist-fraction*, as well as by the *ktourist*-related STi. This result evidences how the combination of general "social distancing" effects (i.e. expressed by *N*-related uncertainty) can be amplified by the specific tourist' occupancy and densities, since *ktourist* describes the effective part of the scenario in which the tourists are generated. Finally, the effect of *initial-distance mean-permanence*, *kresident* and *movement at breaks* appears negligible (STi<0.05).

Figure 3 shows the STi and the SFi results for the second simulation set, that considers *initial-distance*, *mean-permanence*, *kresident* and *movement at breaks* as constant parameters (STi<0.05). Considering a simplified input of variables, the trend of Figure 2 does not substantially change.

Finally, in both the simulation sets, the sum of SFi is less than 1, thus confirming that limited interactions between input factors exist (Saltelli et al., 2007).

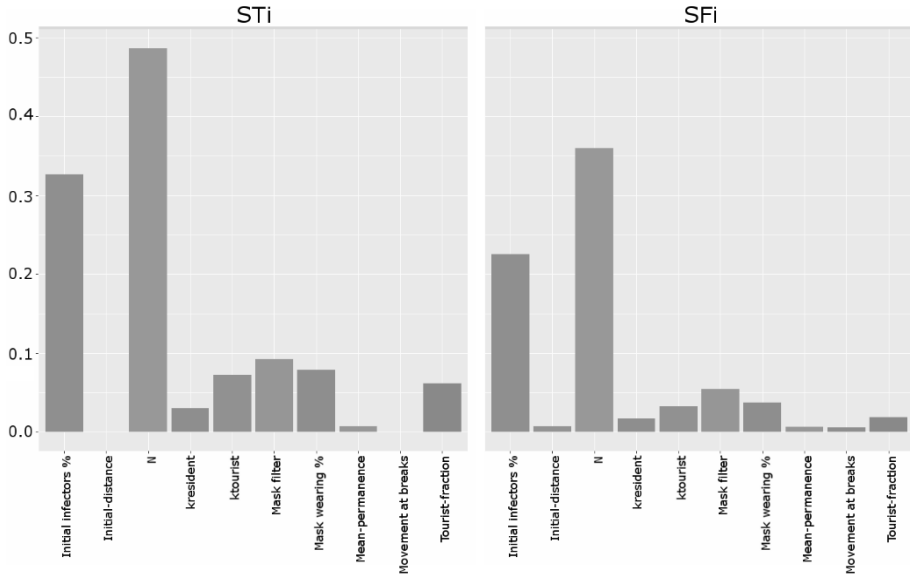

*Figure 2. First simulation set: first-order sensitivity indices (STi - left) and total order sensitivity indices (SFi - right) for the considered parameters.*

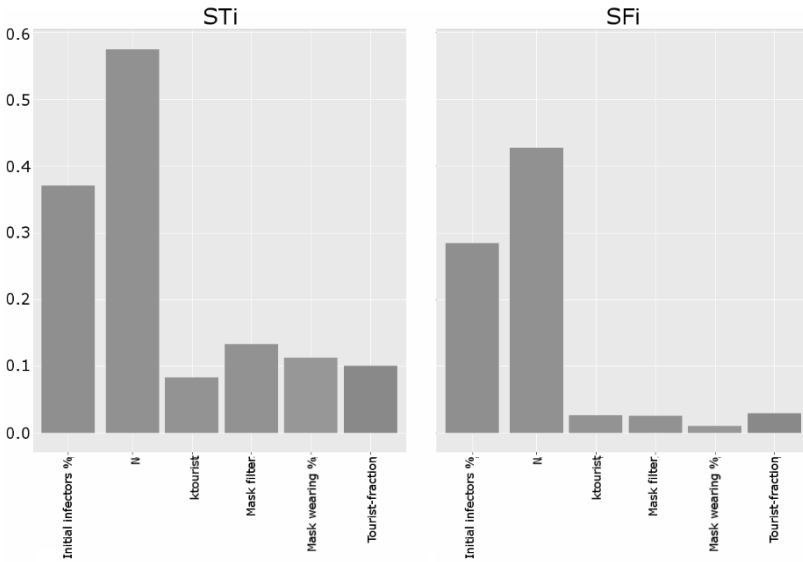

*Figure 3. Second simulation set: first-order sensitivity indices (STi - left) and total order sensitivity indices (SFi - right) for the considered parameters.*

### 3.2. Simulation scenario results

#### 3.2.1. Analysis of the whole simulation sample

Regardless of simulated population and of the *initial infector %*, the use of facial masks by the simulated agents can sensibly reduce the virus spreading, especially in case of the higher *mask filter* values and for widespread adoption of this risk-mitigation measure (higher *mask wearing %*). Previous works suggested a similar impact (D'Orazio et al., 2020; Zhai, 2020). Figure 10Figure 4 shows the distribution of *dI* values for the different *mask wearing %*mask filter* classes, by tracing the *dI* acceptability thresholds of 5% and 25%. The acceptability thresholds are guaranteed in the 75% of the simulations by involving *mask wearing %*mask filter ≥0.50* (e.g. adoption of surgical masks by the whole population) for *dI*=25% and ≥*0.80* (e.g. adoption of FFP1 by the whole population) for *dI*=5%. Such a result is remarked in Figure 5, which considers the *mask filter* classes and traces *mask wearing %* in percentage terms.

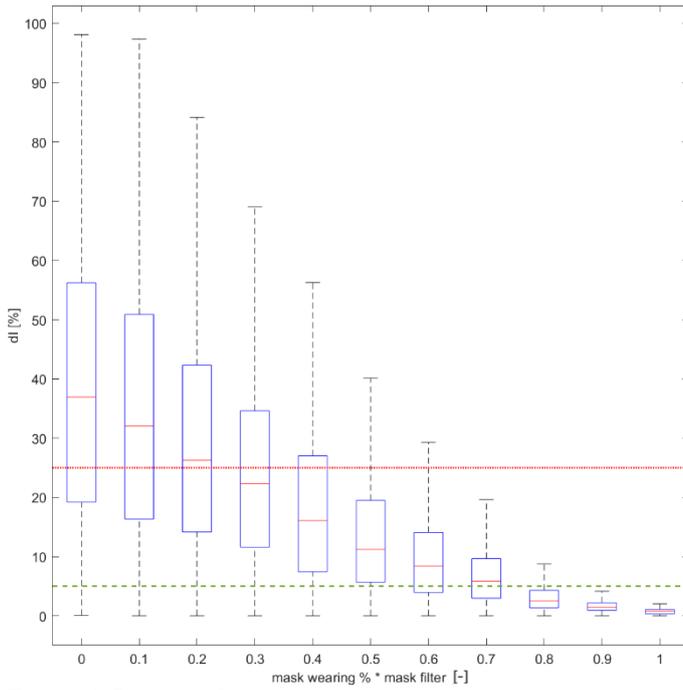

*Figure 4. Boxplot dI values distribution at the last simulation step for the whole sample, concerning the effects related to mask (mask wearing %*mask filter). dI acceptable thresholds are defined at dI=5% (dashed green line) and 25% (continuous red line).*

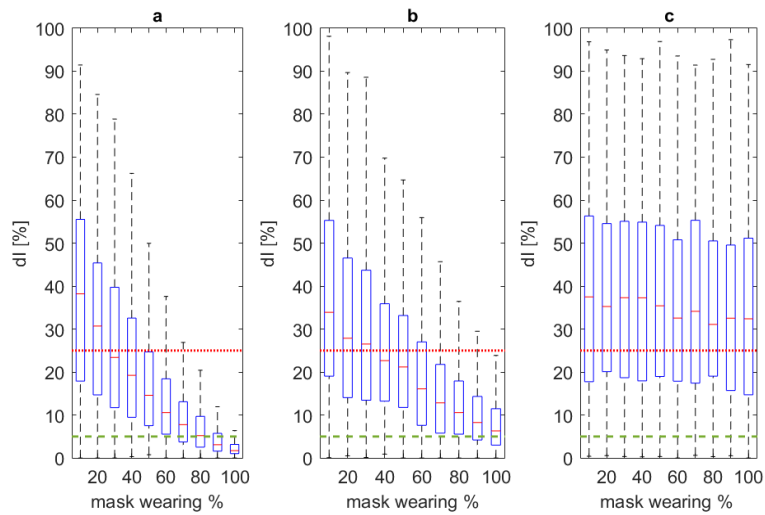

*Figure 5. Boxplot dI values distribution at the last simulation step for the whole sample, concerning the effects related to mask wearing % (expressed in %) by considering Section 2.3 classification: a) FFPk masks; b) surgical mask; c) non-standard protection (0 to 0.25). dI acceptable thresholds are defined at dI=5% (dashed green line) and 25% (continuous red line).*

*Figure 6* suggests that the limitation of the hosted population capacity for the considered urban area limitedly guarantee acceptable solutions in terms of *dI* value, regardless of the adopted additional measures, by confirming results for closed environment application of the model (D'Orazio et al., 2020). In particular, the limitation to 30% of the maximum population (*Docc*=0.3) could support the limitation of *dI≤*25% for the 75% of simulated cases. The combination between such "social distancing"-related measure and the use of facial masks can boost the positive effects, as shown by *Figure 7*. It is worthy of notice that:

- for *Docc≤*0.25, no additional facial masks-related strategies are essentially needed in the 75% of cases if considering *dI*=25%, while surgical masks seem to be enough if considering *dI*=5% (*Figure 7*-A). Nevertheless, such a solution can strongly affect economic aspects in the urban areas, since the limitation to the number of tourists is extremely severe;
- the implementation of surgical masks by the whole population can lead to acceptable solutions in 75% of cases if considering *dI*=25% also for the higher *Docc* values (up to the maximum capacity - *Figure 7*-D);

- the application of FFPk masks by the population is necessary if considering the acceptability threshold at *dI*=5% for *Docc*>0.25 (*Figure 7*-B, *Figure 7*-C, *Figure 7*-D);
- the variation of *dI* distribution (e.g. distance between the maximum and minimum values) is reduced for lower *Docc* values because the possibility to maintain "social distancing" strategies is higher, as well as the possibility to stochastically have additional contacts with infectors, due to the wide urban area.

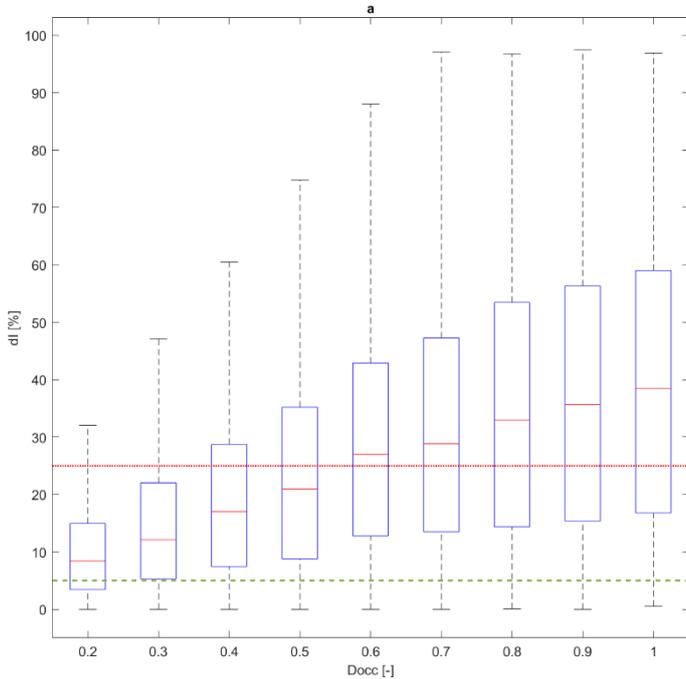

*Figure 6. Boxplot dI values distribution at the last simulation step for the whole sample, with respect to the effect of Docc values discretized by 0.1. dI acceptable thresholds are defined at dI=5% (dashed green line) and 25% (continuous red line).*

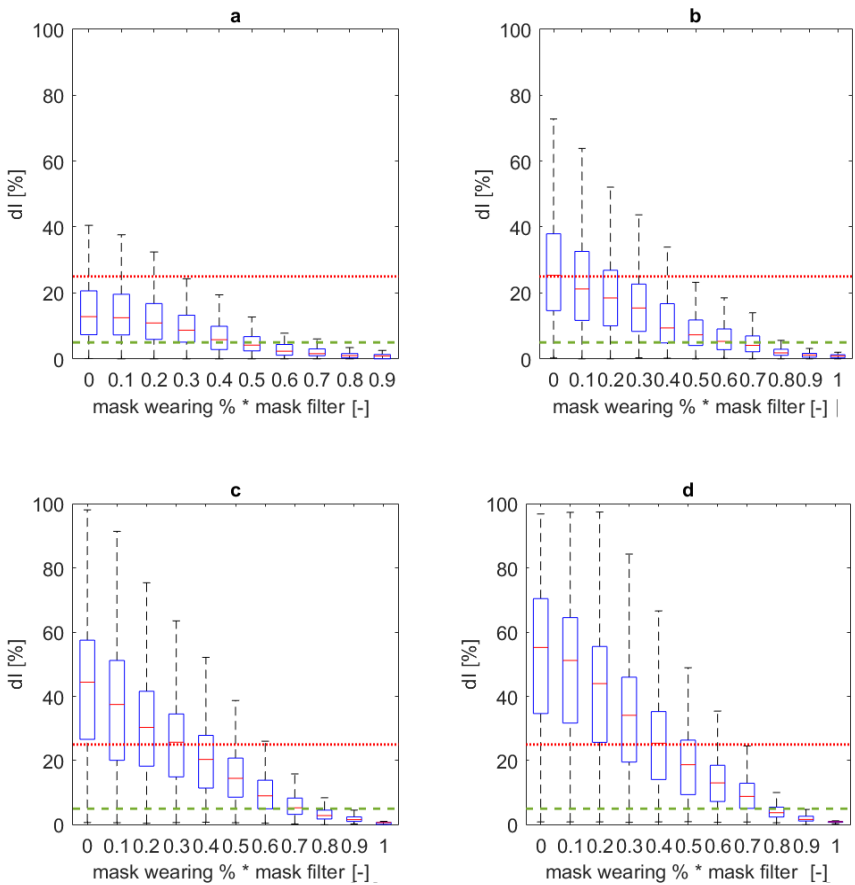

*Figure 7. Boxplot dI values distribution at the last simulation step for the whole sample, with respect to the effects of different density classes: a) Docc≤0.25; b) 0.25<Docc≤0.5; c) 0.5<Docc≤0.75; d) 0.75<Docc≤1.00 . Values are traced according to the overall mask effect. dI acceptable thresholds are defined at dI=5% (dashed green line) and 25% (continuous red line).*

Finally, *Figure 8* traces the probability of maintaining *dI* under the acceptability thresholds in respect of the *initial infector %*, regardless of the implemented risk-mitigation strategies. In general terms, current lower common values in the *initial infector %* (0.14 and 0.30) generally have a significant probability level, especially for *dI*=5%. Meanwhile, *Figure 9* traces the probability distribution of the *dINF* values depending on different thresholds in the initial infector % (compare to Section 2.3). The probability that the contagion could not worsen at the end of the simulation time is higher if considering the current lower common values in the *initial infector %* (0.14 and 0.30), as displayed by *Figure 9*-B (the probability values are significantly higher for *dINF*≥0 in respect to negative values), regardless of the implementation of risk-mitigation strategies. This simulation outcome can be essentially related to the lower probability to have close contacts with an infector within the urban environment during the simulation time.

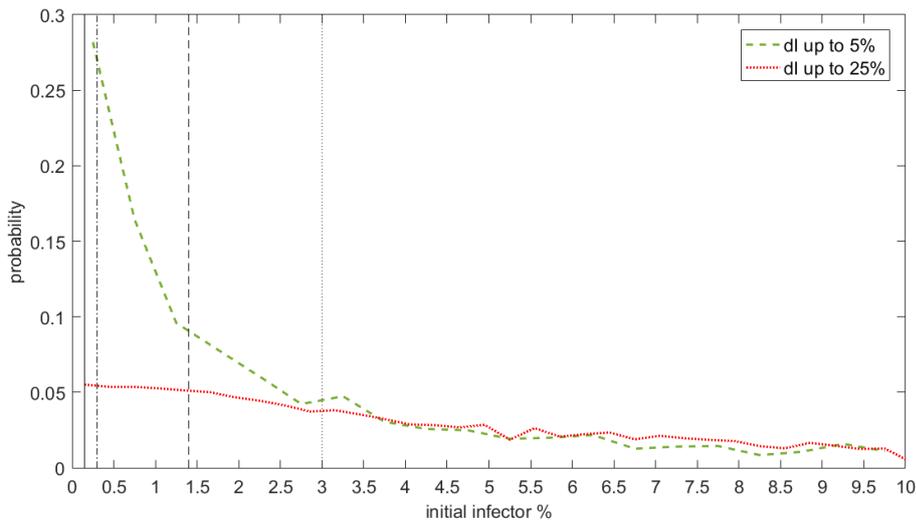

*Figure 8. Probability estimation of initial infector % values that can lead to dI values under the two acceptability thresholds: dI≤5% (dashed green line) and dI≤25% (continuous red line). Significant limits for initial infector % are evidenced according to Section 2.3.*

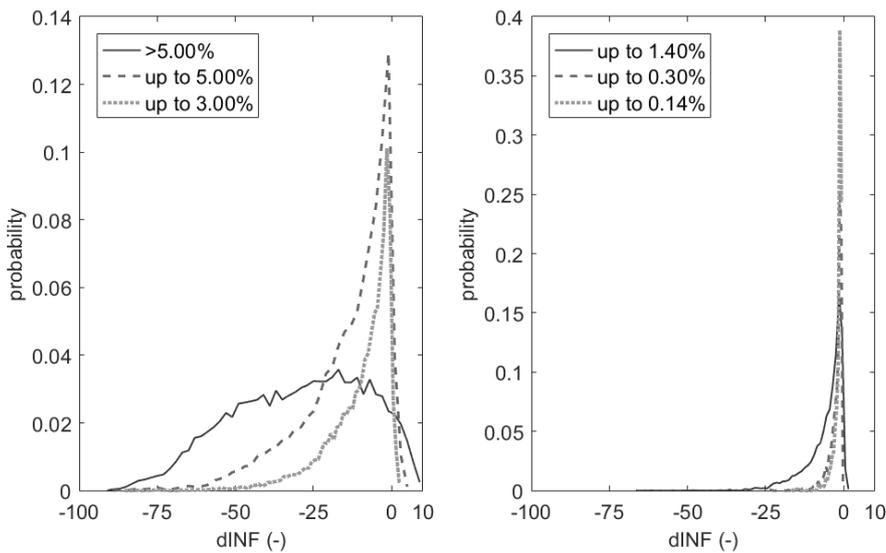

*Figure 9. Probability estimation of dINF according to the significant initial infector % values according to Section 2.3, by distinguishing: a) critical conditions for the "second phase"; b) common infectors' percentages according to the data from the early May 2020.*

### 3.2.2. Analysis considering the maximum population capacity

This section focuses on the maximum capacity conditions for the touristic area (*0.75<Docc≤1.00*). Figure 10 shows the distribution of *dI* values depending on the specific *mask filter* classes and depending on the implementation level in terms of *mask wearing %,* regardless of the *initial infector %*. The contagion spreading could be limited by implementing FFPk masks by the whole population (Figure 10-A), thus leading to *dI<5%*. The application of surgical masks for a wide number of the population seems to bring mitigation effects only if considering *dI=25%* (Figure 10-B). Less than the 25% of the cases involving non-standard protection solutions leads to *dI< 25%*, essentially because of the possibility that wide urban spaces can still guarantee the adoption of limited "social distancing" behaviours. On the contrary, previous simulations relating to single buildings/closed environment underlines that no cases under this threshold could exist, essentially because of the higher effective density inside the building rooms (D'Orazio et al., 2020). These results substantiate the general considerations of Figure 5-A.

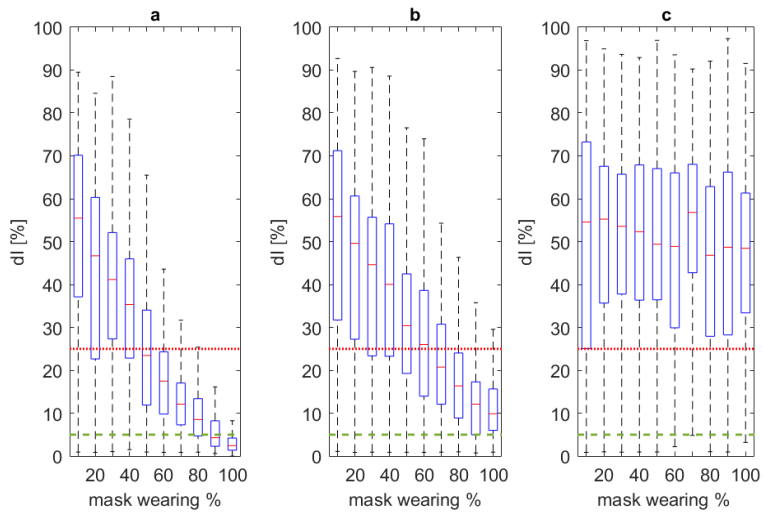

*Figure 10. Boxplot dI values distribution at the last simulation step for Docc between 0.75 and 1.00, with respect to the effects related to mask wearing % (expressed in %) and by considering the Section 2.3 classification: a) FFPk masks; b) surgical mask; c) non-standard protection (0 to 0.25). dI acceptable thresholds are defined at dI=5% (dashed green line) and 25% (continuous red line).*

Figure 11 traces the relation between the *initial infector %* (different panels), the adopted *mask filter* class and *dI* distribution, regardless of the *mask wearing %.* The current main conditions in the *initial infector %* (Figure 11-A, Figure 11-B and Figure 11-C) can generally lead to the possibility to control the contagion spreading by using surgical masks. Nevertheless, the acceptability threshold of *dI=25%* is generally satisfied in the 75% of the cases up to an *initial infector % equal to 5%* (Figure 11-D and Figure 11-E), while FFPk masks generally have a similar impact for all the current main conditions in the *initial infector %.* Values of *initial infector %* over the limit of 5% generally implies critical conditions for the contagion spreading (Figure 11-F). This result confirms the importance of possible "infection-by-chance" for the lowest *initial infector %*, as additionally remarked by the probability estimation for the *dINF* values shown by Figure 12.

In view of the above, in case of surgical of FFPk masks implementations, the median of *dI* distributions for *initial infectors % ≤*0.14% is under the 5% acceptability threshold of *dI*, while the median values up to *initial infectors % ≤*3% are under the 25% (see Figure 11). Nevertheless, since this analysis does not consider the impact of *mask wearing %,* further insights are needed. Hence, data are analysed to provide a quick and simple approach to support decision-makers in the evaluation of the effectiveness of *facial mask* implementation levels by the *mask wearing %* of the population, considering the *initial infectors %.* Data on the *mask wearing %* are simply aggregated by considering the hosted population quartiles to have a look at a glance. Results allow to better stress the general effects of the (coarsely approximated) minimum *mask wearing %* on *dI*, in a rapid application perspective to the case study.

Figure 13 and Figure 14 respectively trace *dI* values for surgical masks and FFPk masks, depending on the *initial infector %* (increasing limits in each panel from A to F) as a function of the *mask wearing %.* Such results confirm the general trends of Figure 11. In particular, current minimum *initial infectors %* values could be managed by implementing surgical masks or FFPk masks for at least the 75% of the population to limit *dI* at 5% (Figure 13-A and Figure 14-A). No significant difference between these two conditions for *initial infectors %≤*0.14% seems to confirm the aforementioned "infection-by-chance" scenario, which gives minor importance to the use of different mask filter. If the *initial infectors*

*%* increases the minimum *mask wearing %* has to increase, too, to guarantee a sustainable *dI*. For *initial infectors %*≤0.30%, surgical masks could guarantee at most *dI* up to 25% for *mask wearing %*≥50% (Figure 13-B), while FFPk could gain *dI*=5% for *mask wearing %*≥75% (Figure 14-B). The use of surgical masks for *mask wearing %*≥75% could guarantee *dI*=25% for *initial infectors %* up to 1.40% (compare Figure 13-C to Figure 13-D and Figure 13-E), while the same *mask wearing %*≥75% could lead to the same result up to 5% if implementing FFPk mask (Figure 14-C, Figure 14-D and Figure 14-E). Although both surgical and FFPk masks can reduce *dI* by increasing the *mask wearing %* up to 75-100%, scenarios related to *initial infectors %* up to 5% are critical for both these protection solutions (Figure 13-F and Figure 14-F).

As a consequence, the implementation of acceptable solutions (from the users' perspective) about facial mask use (e.g. surgical masks) could have significant impacts to limit the contagion in touristic areas during the "second phase" of the COVID-19 emergency by considering the current statistics on active cases. The rule of such protective solution is also in line with previous works suggestions (Howard et al., 2020; Zhai, 2020). Meanwhile, results suggest that the tourists' capacity could be increased towards upper-limit conditions, by ensuring proper economic effects on the communities of these urban areas.

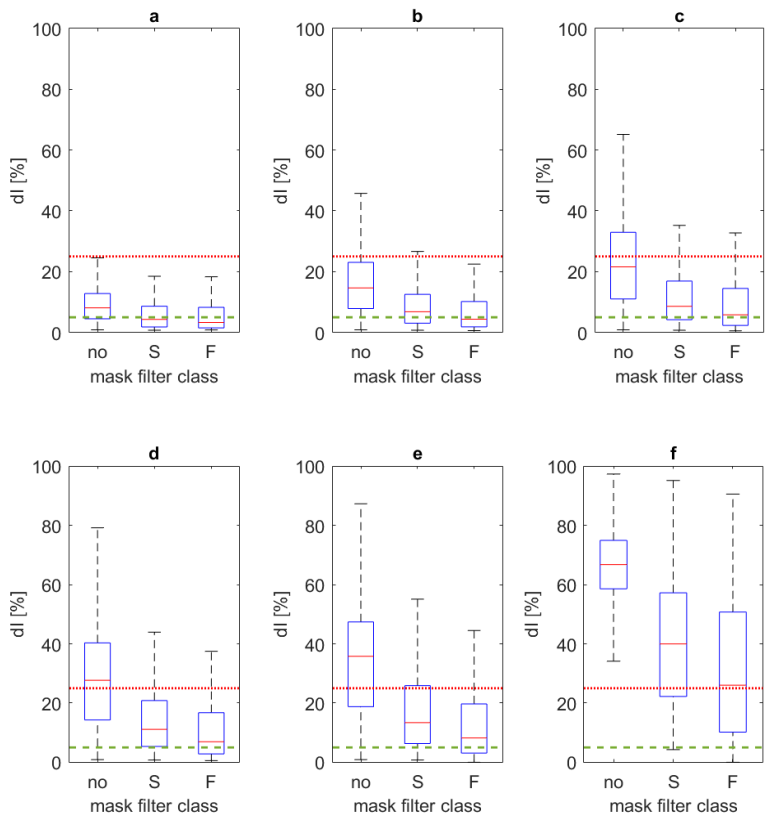

*Figure 11. Boxplot dI values distribution at the last simulation step for Docc between 0.75 and 1.00, with respect to the mask filter classes defined in Section 2.3 for different initial infector % limits: a) up to 0.14%; b) up to 0.30%; c) up to 1.40%; d) up to 3.00%; e) up to 5.00%; f) over 5.00%. Mask filter classes are identified by: "no" for non-standard protection (0 to 0.25); "S" for surgical masks; "F" for FFPk masks. dI acceptable thresholds are defined at dI=5% (dashed green line) and 25% (continuous red line).*

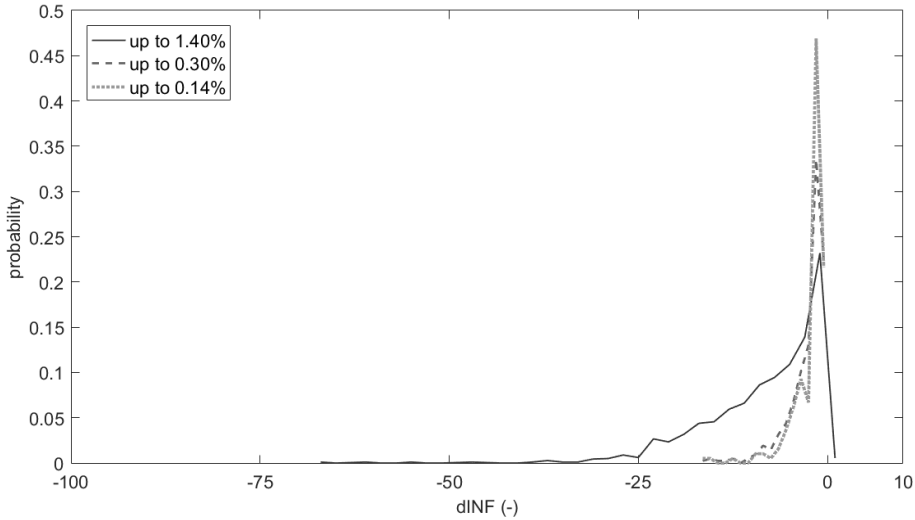

*Figure 12. Probability estimation of dINF according to the significant initial infector % values according to Section 2.3, by distinguishing the common infectors' percentages according to the data from early May 2020.*

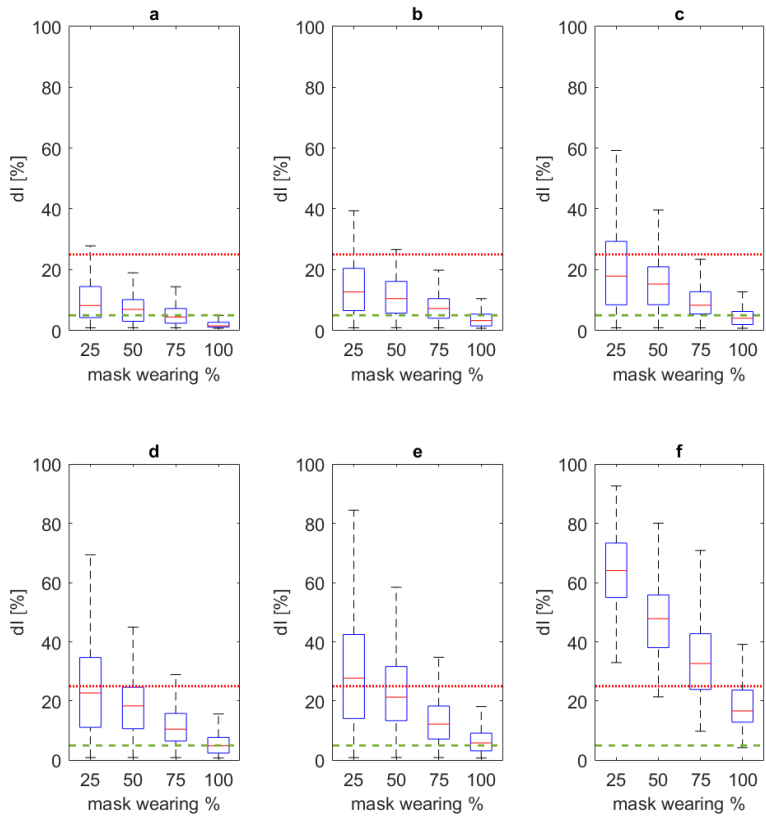

*Figure 13. Boxplot dI values distribution at the last simulation step for Docc between 0.75 and 1.00, with respect to the implementation of surgical masks and to the mask wearing % classes (expressed in %), for different initial infector % limits: a) up to 0.14%; b) up to 0.30%; c) up to 1.40%; d) up to 3.00%; e) up to 5.00%; f) over 5.00%.*

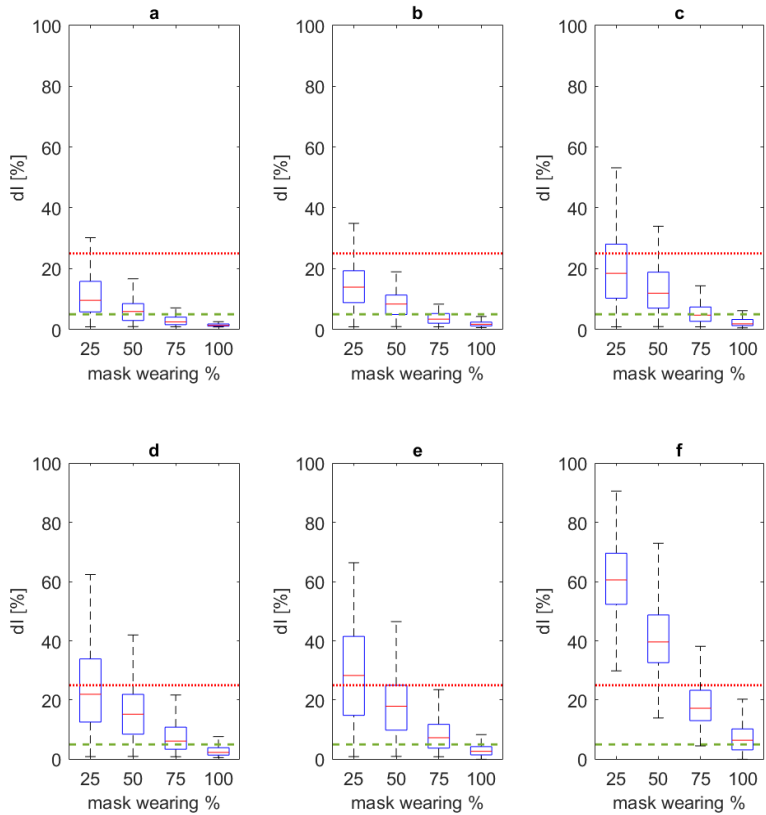

*Figure 14. Boxplot dI values distribution at the last simulation step for Docc between 0.75 and 1.00, with respect to the implementation of FFPk masks and to the mask wearing % classes (expressed in %), for different initial infector % limits: a) up to 0.14%; b) up to 0.30%; c) up to 1.40%; d) up to 3.00%; e) up to 5.00%; f) over 5.00%.*

### 3.3. Discussion about the return to "business as usual" in touristic cities and of model application

Results show how the selection of risk-mitigation strategies in the context of the return to "business as usual" in touristic cities should take into account the current conditions of the virus spreading within the population, so as to make them more effective and acceptable. On one side, combined scenarios in which facial masks are used by the population and a limitation of the occupants' capacity (towards "social distancing"-related strategies) seem to have the highest impact on the possibility to limit the virus spreading, thus confirming previous researches (D'Orazio et al., 2020; Feng et al., 2020; Ferguson et al., 2020; Howard et al., 2020; Zhai, 2020). Nevertheless, their impact is not the same for all the *initial infector %* conditions, as additional remarked by *Figure 15* and *Figure 16*.

*Figure 15* and *Figure 16* respectively trace the simulation *mask filter-Docc* pairs which can lead to *dI* conditions under the lower acceptability threshold (5%), divided by the initial infector % classes defined in Section 2.3, and by additionally tracking the *mask wearing %* values. The proposed overview on the *initial infector %* classes of *Figure 15* and *Figure 16* could be also viewed as a set of imaginable future scenarios due to the possibility to contain or not the infection risk at the starting of the "second phase". Furthermore, since these representations are based on *mask filter-Docc* pairs and do not directly include the *mask wearing %* as main prediction input for *dI*-related limit conditions, the trends of *Figure 15* and *Figure 16* could be also adopted when the users wear facial masks with different *mask filter* values. In this case, the reference *mask filter* can be conservatively considered equal to the lower implemented one.

*Figure 15* refers to the main current conditions in terms of *initial infector %* limits (up to 1.40%), while *Figure 16* traces the results for critical conditions (over 1.40%). *Figure 15* generally evidences a poor level of correlations between the *mask filter-Docc* pairs in case of initial infector % values lower than the current national maximum one (1.40%, compare to Section 2.3). These scattered pairs seem to suggest that current conditions are essentially affected by "infection-by-chance" while moving in the urban scenario. The limitation of the tourists' capacity could not guarantee by itself the acceptability threshold, while the significant economic impact will appear because of the reduction in the number of users within the built environment and its facilities (Gössling et al., 2020).

Additional comparisons between these results and Section 3.2.2 outcomes for the maximum capability conditions follow. According to Figure 12 data on *dINF* statistics provide additional suggestions, the cumulative probability that the contagion will not increase in such an "infection-by-chance" scenario is equal to about 0.5 for *initial infector %*≤0.14% (average national data for Italy at the early May 2020). This percentage seems not to be affected by the *Docc* conditions (compare to *Figure 9*-B), thus confirming the scattered *mask filter-Docc* pairs in *Figure 15* (e.g. *Figure 15*-A for *initial infector %*≤0.14%).

Nevertheless, the implementation of facial masks solution could guarantee the access to the urban areas while having a higher impact on the *dI* values (also compare to Figure 11-A for *initial infector %*≤0.14%) (D'Orazio et al., 2020; Fang et al., 2020; Howard et al., 2020). Hence, the current conditions in the number of *initial infector %* seem to underline the importance of such a non-pharmaceutical intervention in view of the return to "business as usual" in touristic cities: surgical masks could be an adequate protection measure also at the maximum *Docc* values, according to Section 3.2.2 results, while having higher comfort levels on the users in respect to FFPk masks. Meanwhile, other strategies aimed at limiting indirect contagion spreading (e.g. indoor spaces/surface disinfection; use of disposable gloves) should be performed to ensure a higher protection level for the visitors and the workers (Hamid et al., 2020; Pradhan et al., 2020)[10].

On the contrary, *Figure 16* shows how critical conditions in the *initial infector %* implies a more significant impact of the combination between *mask filter* and *Docc* measures, thus confirming previous simulation outcomes for the closed environment (D'Orazio et al., 2020). For *initial infector %*>1.40%, the possibility to have close contacts with an infector is significantly higher in respect to inferior limit conditions, as for high-density indoor scenarios. *Figure 16*-A and *Figure 16*-B qualitatively underline how the occupants' capacity for the urban areas in case of poor facial masks-based solutions should be reduced to the 20 to 40% of the maximum one, while the implementation of surgical could be not enough for *initial infector %* >5%.

These results are confirmed by considering the interpolations of maximum increasing *mask filter-Docc* pairs in *Figure 16* for each considered *initial infector %* range. The interpolations are provided according to a power-based regression approach ($ax^b+c$), as shown by *Table 3*. According to *Figure 16*, the higher the *initial infector %*, the more restrictive the limit for minimum *mask filter-Docc* combination. It is worth notice that such regressions are limited to the considered ideal case study and are not generalizable to other contexts as an operative tool for decision-makers. Nevertheless, they offer an estimation of the upper boundary limit in *mask filter-Docc* combination that should not be overcome (no admitted solution seems to exist over the curve). This methodology could be applied to other case studies to obtain the obtained curves on a "case-by-case" approach, and to finally compare the general trends towards common and simplified rules for decision-makers.

Given the above, decision-makers should be aware of the following main key factors for the sustainability of non-pharmaceutical interventions against COVID-19 spreading in this "second phase":

1. assessing the effective local conditions in terms of infection spreading among the population (i.e. monitoring campaign on active cases), to avoid as far as possible the adoption of limitation to tourists' capacity which: a) could not have the desired effect on the contagion for current main lower percentages of active cases; b) will can depress the tourist-related economic and social issues at the urban scale;

2. promoting the implementation of facial masks for both residents and tourists, by starting from the adoption of comfortable solutions (such as surgical masks);

3. organizing activities over urban spaces and holiday time towards the creation of "widespread" fruition models of the tourists' attractions, to reduce close effects of overcrowding conditions (also by implementing, e.g., reservation-based access to areas and activities). Such choices should be discussed with tourist services stakeholders for tracing general scheduling based on acceptable economic limits.

---

[10] https://www.ecdc.europa.eu/en/publications-data/disinfection-environments-covid-19 (last access: 15/05/2020)

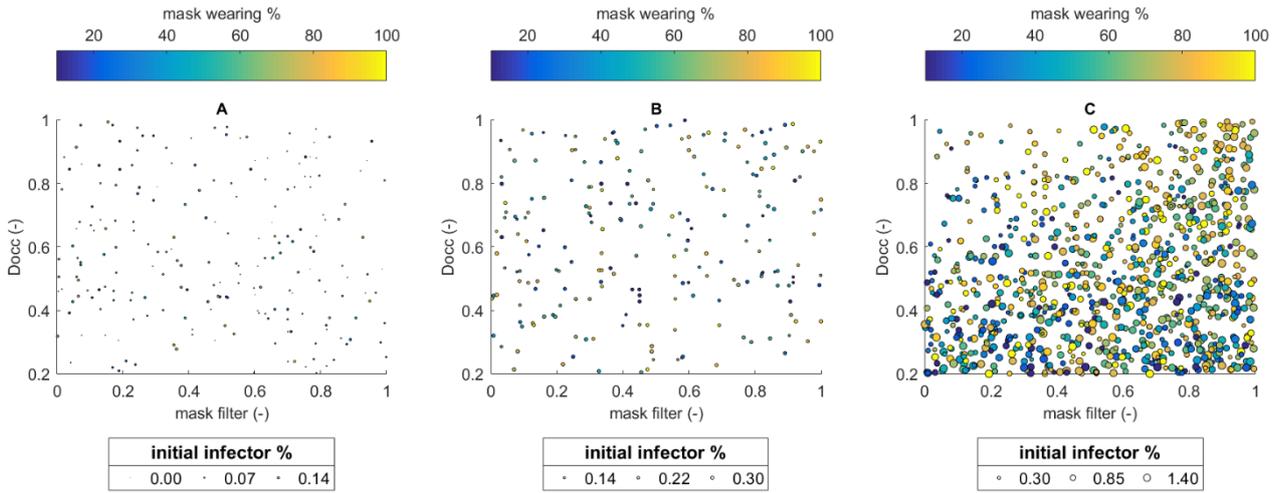

*Figure 15. Mask filter-Docc correlation for all the pairs related to dI≤5% depending on the initial infector percentages classes: a) up to 0.14%; b) from 0.14% to 0.30%; c) from 0.30% to 1.40%. The pairs' colour is related to the mask wearing % (colour bar on the top), while the dot size depends on the initial infector % value (circles inside the legend on the bottom).*

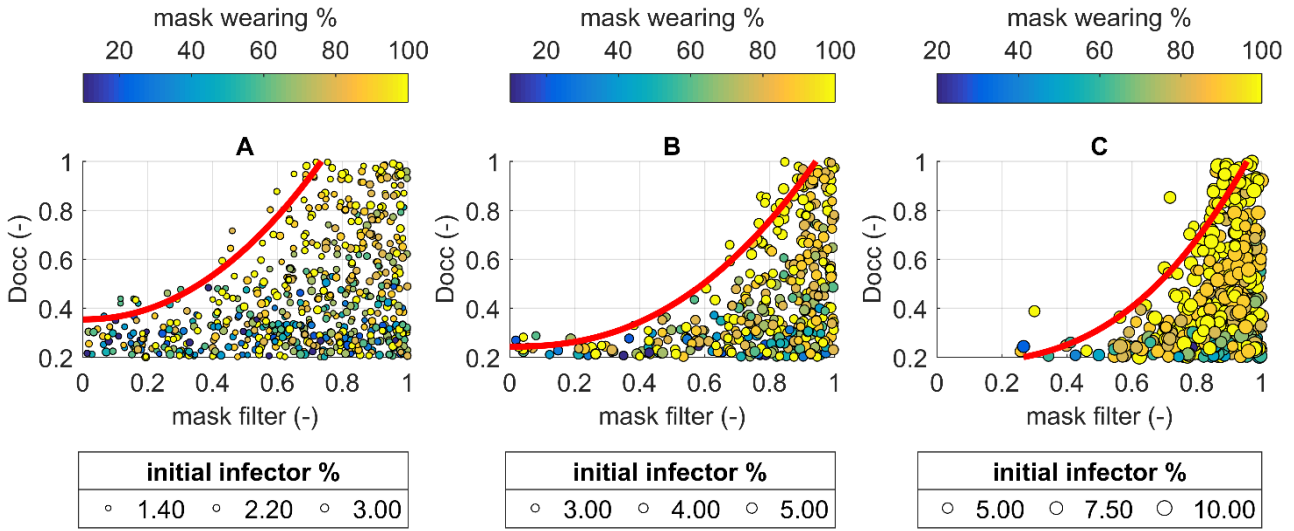

*Figure 16. Mask filter-Docc correlation for all the pairs related to dI≤5% depending on the initial infector percentages classes: a) from 1.40% to 3.00%; b) from 3.00% to 5.00%; c) over 5.00% to maximum value (10.00%). Regression curves for each correlation are shown by the red lines (form: ax^b+c; regression data in Table 3). The pairs' colour is related to the mask wearing % (colour bar on the top) while the dot size depends on the initial infector % value (circles inside the legend on the bottom).*

| initial infector % | *from 1.40% to 3.00* | *from 3.00% to 5.00%* | *over 5.00% to maximum value (10.00%)* |
|---|---|---|---|
| **Equation: y=ax^b+c** | $y=1.29x^{2.09}+0.35$ | $y=0.87x^{2.36}+0.24$ | $y=0.93x^{2.76}+0.17$ |
| **x domain limits** | *0.00 to 1.00* | *0.00 to 1.00* | *0.27 to 1.00* |
| **R²** | *0.93* | *0.94* | *0.86* |

*Table 3. Equation (form: ax^b+c) for Mask filter-Docc correlation curves in limit conditions as drawn by Figure 16. In the equations, x is mask filter, while y is Docc. The limit of the mask filter values domain in which the equation could be applied, as well as the R² are evidenced.*

In this sense, this work underlines how simulation tools could be a significant support to increase decision markers awareness towards ones of the most significant variables affecting the man-man and man-environment interaction in a pandemic. Future activities on the model could involve further calibration task according to future available experimental data in significant urban contexts, to additionally evidence how the differences about modes of transmission and built environment layout/use (e.g. indoor/outdoor; specific activities carried out by the tourist; specific building systems; scheduling of the activities to trace the dependencies from differences in the exposure timing) could affect the overall results (Prussin et al., 2020; Ronchi and Lovreglio, 2020; Zhang et al., 2018).

## 4. Conclusions

After the lockdown phase for the COVID-19 emergency, the return to "business as usual" in touristic urban areas is seriously affected by the possibility to the control contagion spreading due to the flows. The limitation of travels towards touristic area will not be acceptable in a "second phase" of the contagion, essentially because of the necessity to avoid an economic and social crisis for the sector and the involved communities. On the contrary, support tools for decision-makers should be developed to define the effective impact of different sustainable and combined non-pharmaceutical interventions in view of the tourist activities restarting. This kind of assessment should involve at least the scale of homogeneous urban areas, to take into account the general behaviours of the users in the built environment.

This paper modifies an existing agent-based model approach to estimate the Coronavirus spreading in a touristic urban context, by including the simulation of people's movements in the urban areas and the effects of non-pharmaceutical strategies (i.e. facial masks use by agents; occupants' capacity control as the main driver to promote "social-distancing"). The model could be both used to evaluate, over time, how many infectors can appear within the urban area and how many visitors can return home being infected. In this study, the attention is focused on the first point, to focus on the effectiveness and sustainability of strategies on the selected area.

The simulator is applied to a significant case study (an idealized part of a touristic coastal city in Italy) to evidence the general impact of input conditions on the infections over time. Results show the model capabilities in predicting the contagion spreading depending on input variables (including the initial percentage of active COVID-19 cases), thus being a tool to improve the decision-makers' awareness about the impact of contagion-mitigation strategies. In particular, results underline how the adoption of social distancing strategies could not have a leading effect on the contagion spreading when the percentage of initial active cases is close to 0, while becomes an effective strategy in case of critical infectors percentages. At lowest occupants' capacity values, for the current percentage conditions in terms of active cases (e.g. 0.3% of the population or lower), the possibility to be infected in the urban area seems to be more connected to stochastic effects of man-man interaction rather than to a systematic spreading of the contagion. On the opposite, facial masks have a prominent effect on the contagion limitation, especially at lower percentages of active cases. The correlation between the facial mask characterization (i.e. filtering) and the "social distancing"-related strategies (i.e. using tourists' capacity limitation) evidence a clear frontier in the possible combination of these solutions, according to the results for the considered case study. For active cases percentage conditions over 1.40%, the higher the percentage of the active cases in the urban area, the more restrictive the minimum acceptable combination between these two non-pharmaceutical solutions.

From this point of view, decision-makers should then evaluate which maximum tourists' capacity could be applied, by including facial masks-based solutions, to allow the restarting of tourism-related economic activities from a sustainability perspective. According to the results for the case study application, wearing surgical masks could be enough to face main current active cases conditions (at early May 2020; active cases of about 0.14 to 0.30% of the overall population) in touristic urban spaces. The application to further case studies could validate such suggestions. Furthermore, correlations on the minimum acceptable combination between facial masks-based and "social distancing"-related strategies could be assessed to define "case-by-case" decision rules, as well as common criteria for touristic urban areas.

Finally, the agent-based modelling approach will allow the introduction of modifications to integrate epidemiological data (i.e. additional modes of virus transmission), built environment configurations (e.g. indoor/outdoor; including layout characterization), visitors' schedule and activities in the urban spaces (e.g. including the fruition of buildings with specific tourist-related intended uses, e.g. cultural buildings and so on).